\documentclass[11pt,a4paper]{article}
\pdfoutput=1                   
\usepackage{jheppub}
\usepackage{mathtools}         
\usepackage{graphicx}
\graphicspath{{./Figs/}}
\usepackage{comment}
\usepackage{latexsym}
\usepackage{xspace}
\usepackage{lscape}
\usepackage{color}
\usepackage{hyperref}
\usepackage{bm}
\usepackage{bbm}
\usepackage{relsize}
\usepackage{tabularx}
\usepackage{multirow}
\usepackage{amsmath}
\usepackage{amssymb}
\usepackage[table]{xcolor}
\usepackage{lipsum}
\usepackage{feynmp-auto}
\usepackage{upgreek}
\usepackage{nicefrac}
\usepackage[titletoc]{appendix}

\usepackage{caption}
\usepackage{subcaption}

\usepackage{cleveref} 
    \crefname{equation}{}{}
    \crefname{align}{}{}
    \crefname{figure}{}{}    
    \crefname{table}{}{}
    \crefname{section}{}{}   
    \crefname{appendix}{}{}
    \crefname{footnote}{}{}
    
    \crefalias{subequation}{equation}

\def\beq{\begin{equation}}
\def\eeq{\end{equation}}
\def\baq{\begin{eqnarray}}
\def\eaq{\end{eqnarray}}

\title{Tachyonic production of dark relics:
classical lattice vs. quantum 2PI in Hartree truncation\\
}

\author[a,b]{Kimmo Kainulainen}
\author[a,b]{Sami Nurmi}
\author[a]{Olli Väisänen}

\affiliation[a]{Department of Physics, PL 35 (YFL), 40014 University of Jyv\"askyl\"a, Finland}
\affiliation[b]{Helsinki Institute of Physics, PL 64, 00014 University of Helsinki, Finland}
\emailAdd{kimmo.kainulainen@jyu.fi}
\emailAdd{olli.j.r.vaisanen@jyu.fi}
\emailAdd{sami.t.nurmi@jyu.fi}

\abstract{We study the out-of-equilibrium production of non-minimally coupled self-interacting scalar dark matter during reheating using classical lattice simulations. The outcomes of the classical simulations are in qualitative agreement with the previous results obtained using the quantum 2PI approach in the Hartree truncation. In particular, the novel non-linear resonance found in the 2PI Hartee study is present also in the classical lattice simulations and can dominate the final dark matter yield. For the parameters considered, the difference in final value of the scalar two-point function between the two approaches is a factor of ${\cal O}(1)$.}


\begin{document}

\maketitle

%
\section{Introduction}
\label{sec:intro}
%

Out-of-equilibrium dynamics of quantum matter in the presence of classical scalar fields is involved in many primordial processes. Important examples include resonant phenomena~\cite{Kofman:1994rk,kofman:1997yn,Greene:1997fu,Braden:2010wd,Berges:2002cz} and tachyonic instabilities~\cite{Calzetta:1989bj,Guth:1985ya,Weinberg:1987vp,Bassett:1997az,Felder:2000hj,Felder:2001kt,Dufaux:2006ee} in reheating, the electroweak baryogenesis~\cite{Cline:2000nw,Kainulainen:2001cn,Kainulainen:2002th,Cline:2013gha,Cline:2020jre,Konstandin:2013caa,Kainulainen:2021oqs}, the leptogenesis mechanism~\cite{Buchmuller:2000nd,Beneke:2010dz,Anisimov:2010dk,Dev:2017trv,DeSimone:2007gkc,Garny:2009qn,Garbrecht:2011aw,Garny:2011hg,Dev:2017wwc,Jukkala:2021sku}, and also many dark matter setups with very weakly coupled fields~\cite{Kolb:1998ki,Chung:1998zb,Garny:2015sjg,Garny:2017kha,Tang:2016vch,Markkanen:2015xuw, Fairbairn:2018bsw, Cembranos:2019qlm}. Resolving the strongly non-linear dynamics encountered in such setups often requires non-perturbative application of field theory methods~\cite{Boyanovsky:1992vi,Boyanovsky:1993pf,Baacke:2001zt,Arrizabalaga:2004iw,Arrizabalaga:2005tf,Kainulainen:2021eki}.  

In this work we study non-linear dynamics in the dark matter scenario proposed in~\cite{Markkanen:2015xuw,Fairbairn:2018bsw}, where a spectator scalar singlet with a non-minimal coupling $\xi R \chi^2$ undergoes tachyonic instability during reheating, when the classical Ricci scalar $R$ oscillates from positive to negative values.  The singlet has no non-gravitational couplings to the visible sector and the produced excitations constitute a dark matter component \cite{Markkanen:2015xuw,Fairbairn:2018bsw}. In~\cite{Kainulainen_2023} the quantum dynamics in the setup was investigated using the 2PI approach~\cite{Cornwall:1974vz,Berges:2004yj} in the Hartree truncation. The results of~\cite{Kainulainen_2023} indicate that the tachyonic instability is followed by a novel transient resonant stage driven by the two-point function $\langle \chi^2\rangle$ in the presence of the self-coupling $\lambda \chi^4$. The resonance can enhance the net particle production by an order of magnitude compared to the semi-analytical estimates of~\cite{Fairbairn:2018bsw}, based on~\cite{Dufaux:2006ee}. The results of~\cite{Kainulainen_2023} indicate that the transient resonance occurs somewhat after the point when the effective mass contribution $\xi R$ falls below $\lambda\langle\chi^2 \rangle$, and its strength depends sensitively on the couplings $\lambda$ and $\xi$, and on the equation of state of the universe during reheating. 

In the current work, we run classical lattice simulations in the setup of~\cite{Markkanen:2015xuw,Fairbairn:2018bsw} and make quantitative comparison with the 2PI Hartee results of~\cite{Kainulainen_2023}. There are two conceptually different effects that can generate differences between the two approaches. First, while~\cite{Kainulainen_2023} investigates quantum evolution starting from vacuum initial conditions, the classical lattice simulation describes on-shell dynamics starting from an initial field configuration with classical plane waves. Second, the Hartree truncation used in~\cite{Kainulainen_2023} does not account for momentum exchanging processes and it is a priori unclear to what extent they could affect the resonant growth. All on-shell effects of the mode mixing induced by tree level-lagrangian are included in the classical lattice simulation. A third aspect is that the sources of numerical error in the lattice simulation and in solving the 2PI equations as implemented in~\cite{Kainulainen_2023} are different and it is not a priori obvious which approach is more efficient in this respect. 

We perform the lattice simulations using the CosmoLattice code~\cite{Figueroa_2021,Figueroa_2023}. CosmoLattice and other lattice codes have already been applied to similar non-minimally coupled scalar setups in~\cite{Figueroa:2021iwm, laverda2023ricci}, but for a reheating equation of state and spectator couplings different from those used in~\cite{Kainulainen_2023}, and direct comparison of the results is therefore not possible. The results obtained in this work confirm that the transient resonance seen in~\cite{Kainulainen_2023} is present also in the classical lattice solution and the net particle production is in a broad agreement with~\cite{Kainulainen_2023}. Details of the resonant stage and the final momentum distribution of the two-point function measured from the classical lattice solutions however differ from the corresponding results in~\cite{Kainulainen_2023}.

 This paper is organized as follows. In section~\cref{sec:model} we briefly specify the setup, in section~\cref{sec:lattice} we discuss relevant aspects of the lattice formulation, and in section~\cref{sec:results} we present the lattice results and compare them against~\cite{Kainulainen_2023}. Lastly, section~\cref{sec:conclusions} presents our conclusions. 

%
\section{The setup}
\label{sec:model}
%

We study the setup of~\cite{Fairbairn:2018bsw} with a non-minimally coupled spectator scalar singlet $\chi$, whose action is given by  
\begin{equation}
\label{eq:action}
    \mathcal S_\chi = \int d^4 x \sqrt{-g} \left[\frac12 (\nabla^\mu \chi)(\nabla_\mu \chi) - \frac12 m_\chi^2 \chi^2 + \frac{\xi}{2} R \chi^2 - \frac{\lambda}{4} \chi^4 \right]~.
\end{equation}
Following~\cite{Kainulainen_2023}, we use the $(+,-,-,-)$ sign convention. We study the dynamics during the reheating stage after inflation, assuming the universe is dominated by a homogeneous inflaton field oscillating in a quadratic potential. The classical equations of motion for the setup read  
\begin{align}
&\ddot \chi + 3 H \dot \chi - \frac{1}{a^2} \nabla^2 \chi + (\xi R + m_\chi^2) \chi + \lambda \chi^3 = 0~, \label{eq:spectator_eom}\\
&\ddot \phi + 3 H \phi + m_\phi^2 = 0~, \label{eq:inflaton_eom}\\
& H = \frac{1}{\sqrt{6} M_P} \left( \dot \phi^2 + m_\phi^2 \phi^2 \right)^{1/2}~, \label{eq:hubble}
\end{align}
and the Ricci scalar is given by 
\beq
\label{eq:ricci}
R = \frac{1}{M^2_P} \left( \dot \phi^2 - 2 m_\phi^2 \phi^2 \right)~.
\eeq

In~\cite{Kainulainen_2023}, the inflaton decay was also accounted for by including an effective decay term $\Gamma\dot{\phi}$ and a radiation fluid in the analysis. The resonant growth of $\langle \chi^2 \rangle $ was found to be strongest for $\Gamma = 0$. Here we will set $\Gamma = 0$ in order to compare the 2PI results of~\cite{Kainulainen_2023} against classical lattice simulations in the non-trivial limit where the resonance is expected to have a maximal impact on the dynamics. In reality, the inflaton of course eventually needs to decay into radiation and the $\Gamma = 0$ case should therefore be physically understood as the limit where the inflaton decay can be neglected in the time-scale of our simulation runs. 

Throughout this work we focus on the parameter range, where the singlet $\chi$ is an energetically subdominant spectator. The energy density of $\chi$ with the non-minimal coupling~\cref{eq:action} is given by 
\beq
\label{eq:rho_chi}
\rho_{\chi} =  \left\langle \frac12 \dot\chi^2 + \frac{1}{2a^2} |\nabla \chi|^2 + \frac12 m_\chi^2 \chi^2  + \frac{\lambda}{4} \chi^4 + 3 \xi(H^2 \chi^2 + 2 H \chi\dot{\chi} -\frac{2}{3 a^2}\nabla\cdot(\chi\nabla\chi)) \right\rangle \!, 
\eeq
where the brackets denote volume averages in the case of a single classical lattice simulation and ensemble averages in the quantum computation, (for which the non-commuting terms also need to be symmetrised). The last total derivative term is irrelevant in our case, as it vanishes both in a spatially isotropic quantum system and in a classical setup with periodic boundary conditions. The condition for $\chi$ being a spectator is that $\rho_{\chi} \ll 3H^2M_{\rm P}^2$ throughout the computation so that its contribution to the Friedmann equation~\cref{eq:hubble} can be neglected.

%
\section{The lattice implementation}
\label{sec:lattice}
%

We perform classical lattice simulation of the spectator field $\chi$ during reheating using a modified version of the CosmoLattice code~\cite{Figueroa_2021,Figueroa_2023}. We first solve separately the homogeneous equations~\cref{eq:inflaton_eom} and~\cref{eq:hubble} for the inflaton and the scale factor. The built-in function used by CosmoLattice for $a(t)$ in the case of a fixed background was then replaced by an interpolant of this solution. The spectator field is evolved in this background using CosmoLattice's leapfrog evolver with the non-minimal coupling to the Ricci scalar in~\cref{eq:spectator_eom} added to the evolution kernel. For a comparison, selected runs were also performed using a fourth-order velocity-verlet evolver, whose output agreed with the leapfrog solutions. The spectra and most averaged quantities of the spectator field solution were extracted using built-in CosmoLattice routines, but as in~\cite{Figueroa:2021iwm} the measured field energy was modified to include the corrections from the non-minimal coupling in equation~\cref{eq:rho_chi}. The evolution is performed in comoving direct space on a cubic grid of $N$ points per dimension. The lattice parameters were chosen so that the comoving infrared cutoff corresponds to the initial Hubble scale, $k_{\rm IR} = H_{\rm i}$ (see the text below for the definition of the initial time $t_{\rm i}$). The lattice size should be chosen so that the linear ultraviolet cutoff $k_{\rm UV}^{\rm lin} = N k_{\rm IR} /2$ is well above the sharp drop-offs seen in the spectra presented in~\cite{Kainulainen_2023}. We find that this is achieved with $N = 512$, though larger lattices were used to test the solutions for resolution dependence. The maximum momentum the lattice can contain is $k_{\rm UV} = \sqrt{3} k_{\rm UV}^{\rm lin}$. The system is evolved in conformal time with the timestep $d\tau = 10^{-4} H_{\rm i}^{-1}$  and the calculation is performed in units of the initial Hubble scale $H_{\rm i}^{-1}$. 

\paragraph{Initial conditions.} 

Following~\cite{Kainulainen_2023}, we initialize the homogeneous inflaton sector with slow roll initial conditions at $\phi = 15 M_{\rm P}$. 
Inflation ends at $\epsilon_{\rm H} = -\dot{H}/H^2 =1$, corresponding to a conformal time $\tau_0$.
We start the lattice computation at the initial time $\tau_{\rm i}$ defined slightly before the end of inflation as the moment when $N_{\rm i} = {\rm ln}(a_0/a_{\rm i}) = 2.485$. 

In~\cite{Kainulainen_2023} the quantum field $\chi$ was initialized in the Bunch Davies vacuum with a vanishing one-point function $\langle \chi(\tau_{\rm i},{\bf k}) \rangle =0$ and with the spectrum of the two point function, $\langle \chi(\tau_{\rm i},{\bf k})\chi(\tau_{\rm i},{\bf k'}) \rangle = (2\pi)^3 \delta({\bf k}+{\bf k'}) P_{\rm i}(k)$, given by 
\begin{equation}
    P_{\rm i}(k) = \frac{\pi}{4 a_{\rm i}^2} e^{-\pi \text{Im}(\nu)}(-\tau_{\rm i}) | H^{(1)}_\nu(-k\tau_{\rm i}) |^2   
    \label{eq:bunchdavies_spectrum}~.
\end{equation}
Here $\tau_{\rm i} = -(a_{\rm i} H_{\rm i})^{-1}$ and $H^{(1)}_\nu$ is the Hankel function of the first kind with the index $\nu^2 = \frac14 - 12(\xi - \frac16)$. Here we study the limit $\xi\gg 1$ and $H_{\rm i}\gg m_{\chi}$ where the effective potential for $\chi$ is initially dominated by the the non-minimal coupling. Therefore the field is initially effectively massive and the index $\nu$ is imaginary. For $\xi = 50$ used in this work and in~\cite{Kainulainen_2023}, the modulus squared of the Hankel function is approximated to three digit precision by $ |H^{(1)}_\nu(x)|^2 \approx e^{\pi \text{Im}(\nu)}(2/\pi)/\sqrt{x^2 + |\nu|^2}$ for all argument values. Therefore, the initial spectrum can simply be approximated by the Minkowski result 
\begin{equation}
    P_{\rm i}(k) = \frac{1}{2 H_{\rm i}a_{\rm i}^2} \frac{1}{\sqrt{(k/H_{\rm i})^2 + a_{\rm i}^2 |\nu|^2}},   
    \label{eq:massive_minkowski_ps}
\end{equation}
which is supported in CosmoLattice by default. 

When the corresponding classical system is simulated on a lattice, the initial field configuration $\chi(\tau_{\rm i},{\bf x})$ is drawn from a distribution which generates the same one- and two-point functions with ensemble averages replaced by volume averages over the lattice. We neglect all higher order connected correlators at the initial time as the non-minimal coupling $\xi R\chi^2$ initially dominates the effective potential. The initial field $\chi(\tau_{\rm i},{\bf x})$ therefore has a Gaussian distribution with zero mean and a spectrum determined by the discretised version of~\cref{eq:massive_minkowski_ps}. We note that CosmoLattice discretises a continuum initial spectrum of the form~\cref{eq:massive_minkowski_ps} by substituting $k^2$ with the square of $k^{j}=n^{j} k_{\rm IR}$, where $n^{j}$ labels the lattice sites and $k_{\rm IR}$ is the smallest comoving momentum on the lattice~\cite{Figueroa_2021}. This does not fully coincide with the actual vacuum spectrum on lattice, (see equation~\cref{eq:lattice_vacuum} below), obtained by using the lattice dispersion relation that follows from discretising the derivative operator. However, we have checked that in our setup the spectrum rapidly evolves to~\cref{eq:lattice_vacuum} well before the onset of tachyonic instability.

\paragraph{Subtracting the vacuum.} 

The results for the two-point function in~\cite{Kainulainen_2023} were given in terms of $\delta \Delta_{\rm F} = \Delta_{\rm F}-\Delta_{{\rm F}0}$, where $\Delta_{\rm F}$ is the finite part of the full two-point function and $\Delta_{{\rm F}0}$ the finite part of the vacuum two-point function, computed for the effective mass solved from the 2PI gap equation and excluding the tachyonic modes. To compare with these results, we need to perform a similar subtraction of the vacuum contribution from the two-point function measured on the lattice. 

The lattice vacuum power spectrum can be obtained from equation~\cref{eq:massive_minkowski_ps} by substituting $k^2$ with  
\begin{equation}
    k^2_{\text{lat}} = \frac{4 L^2}{\pi^2} \sum_i \sin^2\left( \frac{\pi k^i}{2 L} \right),
\end{equation}
which is the Fourier space representation of $-\nabla^2$ for the symmetric definition of the lattice derivative operator~\cite{Figueroa_2021}. Here $k^i$ are the Cartesian components of the comoving momentum and $L = N k_{\rm IR}/2$ is half the comoving length of the cubic lattice box side.
In addition,  we must restrict~\cref{eq:massive_minkowski_ps} to only include modes which fit on the lattice and exclude eventual tachyonic modes as was done in~\cite{Kainulainen_2023}. The resulting angle-averaged vacuum power spectrum is 
\begin{equation}
    P_{\text{vac}}(k) = \int \frac{d\Omega}{4\pi} \frac{1}{H} \frac{1}{2a^2} \frac{\Theta(k^2_{\text{lat}} + M^2_{\text{eff}})}{\sqrt{k^2_{\text{lat}}/H^2 + M_{\text{eff}}^2/H^2}} \prod_i \Theta(L - |k^i|),   
    \label{eq:lattice_vacuum}
\end{equation}
where $\Theta$ denotes a step function. The first step function cuts out tachyonic modes when $M_{\text{eff}}^2<0$ and the latter product of three step functions ensures that only modes that fit on the cubic lattice are included. 

We define the effective mass $M_{\text{eff}}^2$ as the solution of  
\begin{align}
    M_{\text{eff}}^2 = &\:\:a^2 m_{\chi}^2 - a^2 \left(\xi - \frac16\right)R + 3 \lambda a^2 \langle \chi \rangle^2 + 3\lambda a^2 \left ( \langle \chi^2 \rangle - \langle \chi^2 \rangle_{\text{vac}} \right) \label{eq:gap_equation}\\
    & + \frac{3\lambda}{16\pi^2}\left[ M_{\text{eff}}^2 \ln\left( \frac{M_{\text{eff}}^2}{a^2 m_{\chi}^2}\right) - M_{\text{eff}}^2 + a^2 m_{\chi}^2  \right]~, \nonumber
\end{align} 
where $\langle \chi^2\rangle $ is the contact limit of the full two-point function measured from the lattice and $\langle \chi^2_{\rm vac}\rangle = \int {\rm d}^3k/(2\pi)^3 P_{\rm vac}(k)$,  with $P_{\rm vac}$ given by equation~\cref{eq:lattice_vacuum}. Equation~\cref{eq:gap_equation}, which can be solved iteratively alongside equation~\cref{eq:lattice_vacuum}, is just the 2PI gap equation of~\cite{Kainulainen_2023} written in terms of lattice quantities and tree-level couplings. 
The gap equation is introduced here only for the purpose of comparing our results with~\cite{Kainulainen_2023}, as to this end we need to define $M_{\text{eff}}^2$ and subtract the vacuum part from the lattice two-point function in same manner as was done in~\cite{Kainulainen_2023}. In particular, the gap equation does not enter the classical lattice equations of motion in any way. However, as we will discuss below, the vacuum spectrum~\cref{eq:lattice_vacuum} with $M_{\text{eff}}^2$ solved from~\cref{eq:gap_equation} matches well with the lattice two-point function before the tachyonic or resonant particle production starts. Finally, following~\cite{Kainulainen_2023} we split $M_{\text{eff}}^2$ into different components as
\begin{align}
\label{eq:Meffcomp}
    &M^2_{R} = - a^2 \left(\xi - \frac16\right)R, \\
    &M^2_{\Delta} = 3\lambda a^2 \left ( \langle \chi^2 \rangle - \langle \chi^2 \rangle_{\text{vac}} \right), \label{eq:Meffcomp2}\\
    &M^2_{\sigma} = M^2_{\text{eff}} - M^2_R - M^2_{\Delta}. \label{eq:Meffcomp3}
\end{align}
We stress that while the effective mass defined by~\cref{eq:gap_equation} allows a sensible definition of the vacuum in the lattice calculation, it differs essentially from its 2PI-counterpart, which, in the 2PI-method, controls the dynamical evolution of the 2-point correlation function and hence of the variance, including the back-reaction from one to another.

%
\section{Lattice results and comparison with the 2PI Hatree approach}
\label{sec:results}
%

We ran the lattice simulation for three setups matching those studied in~\cite{Kainulainen_2023}. The inflaton mass and the initial inflaton field value were chosen to be $m_{\phi} = 1.5 \times 10^{-13}$ GeV and $\phi = 15 M_P$, and the slow roll initial conditions were used for $\dot{\phi}$. The lattice simulations were initialized at $N_{\rm i}=2.485$ e-folds before the end of inflation (corresponding to the inflaton value $\phi = \sqrt{2} M_P$) and we set $a_{\rm i} = 1$. In the following we denote by $a_0 \approx 12.00$ the scale factor at the end of inflation defined by $\epsilon_{\rm H} = -\dot H/H^2 = 1$. We set $\xi = 50$ in all of our simulations and vary the self-coupling $\lambda$. Unless noted otherwise, the runs were performed with $N = 512$ and a comoving computation box size of $H_{\rm{i}}^{-1}$. 

\paragraph{Variances, effective masses and energies.} 

\begin{figure}[t]
     \centering
     \includegraphics[width=0.95\linewidth]{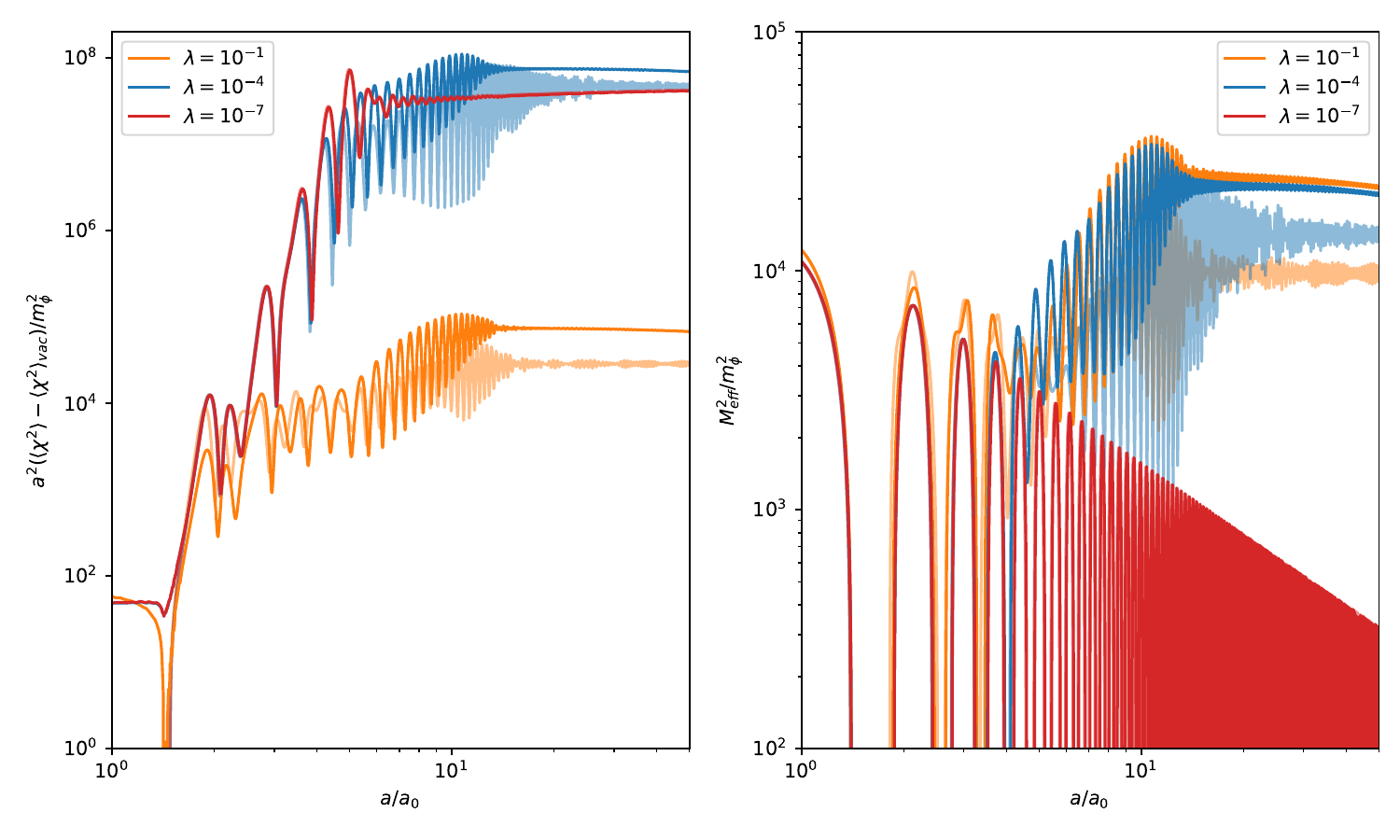} 
        \caption{(Left panel.) The measured lattice variances (solid lines) and the corresponding 2PI solutions reproduced from~\cite{Kainulainen_2023} (transparent lines) for runs with $\lambda = 10^{-1}$, $\lambda = 10^{-4}$ and $\lambda = 10^{-7}$. The vacuum has been subtracted off from all lattice variances, and the 2PI-solution for $\lambda = 10^{-7}$ overlaps with the lattice results. (Right panel.) The lattice results for the effective mass calculated using equation \cref{eq:gap_equation} (solid lines) compared with their 2PI equivalents (transparent lines).}
        \label{fig:varentsiances_M2eff}
\end{figure}

\begin{figure}[h]
     \centering
     \includegraphics[width=0.95\linewidth]{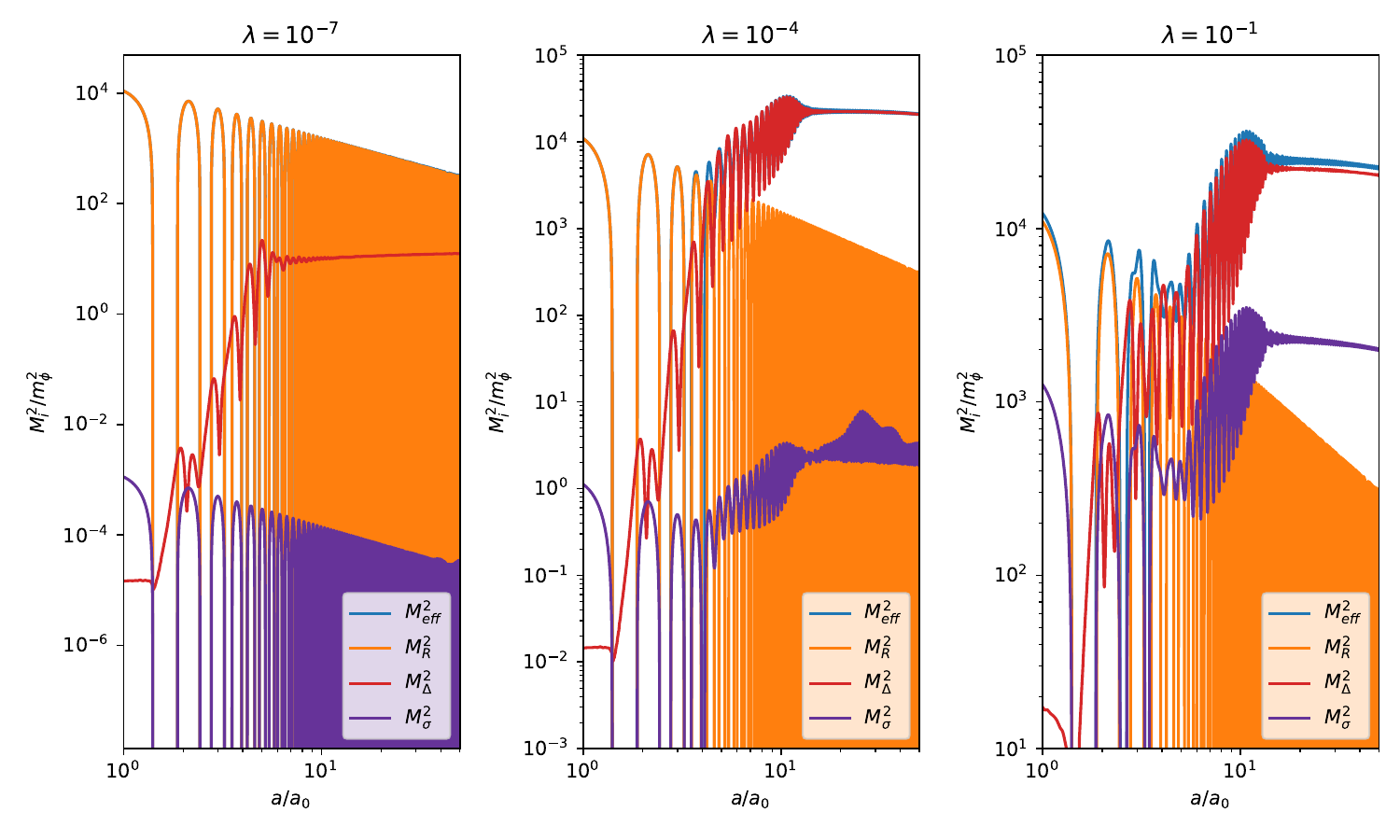} 
        \caption{The components of the effective mass for runs with $\lambda = 10^{-1}$, $\lambda = 10^{-4}$ and $\lambda = 10^{-7}$ as defined in equations \cref{eq:Meffcomp,eq:Meffcomp2,eq:Meffcomp3}.}
        \label{fig:variances_M2eff}
\end{figure}

The left panel of figure~\cref{fig:varentsiances_M2eff} shows the lattice results for the variances $\langle \chi^2\rangle -\langle\chi^2\rangle_{\rm vac}$, where the subtracted vacuum part 
is determined by equation (\ref{eq:lattice_vacuum}). The right panel shows the corresponding squared effective masses solved using the gap equation~\cref{eq:gap_equation}. For comparison, we have also plotted the corresponding 2PI Hartee results of~\cite{Kainulainen_2023} with transparent lines. The lattice results for the effective mass components, defined in equation (\ref{eq:Meffcomp}), are shown in figure~\ref{fig:variances_M2eff}. 

The initial growth of the variance is driven by tachyonic periods where the effective mass becomes imaginary. The end of the tachyonic growth depends on the self-coupling: larger values of $\lambda$ correspond to earlier onset of non-linear dynamics which shuts off the tachyonic growth. For $\lambda = 10^{-7}$ the non-linear region is never reached. Instead, the growth ends as the effective mass decreases and the tachyonic windows become increasingly narrow, as discussed in~\cite{Kainulainen_2023}. For $\lambda = 10^{-4}$ and $\lambda = 10^{-1}$, the tachyonic stage ends when the effective mass contribution from the two-point function, $M_{\Delta}^2$ becomes comparable to $M_{R}^2$. After this the solutions experience a transient resonance during which the lattice results for $\langle \chi^2\rangle -\langle\chi^2\rangle_{\rm vac}$ grow by factors of 5 and 10, respectively. For $\lambda = 10^{-1}$ there is a short period of non-linear oscillations between the end of the tachyonic stage and the onset of the resonance.  After the end of the resonance, the variances redshift close to a law $a^{-2}$. Beyond $a/a_0 \approx 30$, our lattice results start to develop numerical resolution related spurious effects which we discuss at the end of this section. 

The lattice results for the variances and the effective masses closely agree with \cite{Kainulainen_2023} in the region where the dynamics is essentially linear: for $\lambda = 10^{-7}$ this is the case throughout the simulation, and for $\lambda = 10^{-4}$ and $\lambda = 10^{-1}$ until the end of the tachyonic growth. The subsequent non-linear evolution for $\lambda = 10^{-4}$ and $\lambda = 10^{-1}$ also share qualitatively similar features between the classical lattice and the quantum 2PI Hartree solutions. In particular, the lattice results confirm the existence of the non-linear transient resonance which was first observed in \cite{Kainulainen_2023} and which, within the 2PI formalism, is clearly seen to be driven by the two-point function $\langle\chi^2\rangle$. There are however diffences in details of the non-linear dynamics. In the lattice results for $\lambda = 10^{-4}$ the resonant-like growth starts almost immediately after the tachyonic stage when the effective mass contribution $M_\Delta^2$ quickly overtakes $M_R^2$. This is contrary to \cite{Kainulainen_2023}, where one sees a plateau of coexistence between $M_R^2$ and $M_\Delta^2$ at $a/a_0 \sim 5$. However, such a plateau is present in the lattice solution for $\lambda = 10^{-1}$, which is qualitatively very similar to the 2PI runs until the onset of the resonance. In both cases the non-linear growth is more efficient on the lattice and the two-point function saturates to slightly larger final values than in~\cite{Kainulainen_2023}. It can also be noted that oscillations of the two-point function decay faster on lattice at the end of the resonance than in the solutions of~\cite{Kainulainen_2023}. 

We expect the differences arise mostly from the lowest order Hartree truncation used in the  2PI study of \cite{Kainulainen_2023}. In particular, the Hartree truncation neglects momentum exchanging scatterings while the lattice simulation contains the full non-linear dynamics at the classical level. Hartree results therefore do not properly capture processes that smear out coherent oscillations through the self coupling. On this basis, one could have expected that the lattice results would show less prominent resonant growth than  the Hartree results. However, as discussed above, and seen in figure~\cref{fig:varentsiances_M2eff}, the situation is actually the opposite at least for the parameter choices studied in this work. Another fundamental difference is that while the classical lattice simulation of course describes only on-shell physics, the 2PI results of \cite{Kainulainen_2023} capture, within the Hartree truncation, the full quantum evolution of the system. 
Also, the dynamical coupling between the effective gap-mass and the 2-point function and hence the variance in the 2PI-approach may play an important role in explaining the difference.\footnote{Indeed, we have checked that replacing the effective mass in 2PI-equations with the interpolated effective mass from a corresponding lattice run leads to an inconsistent 2PI-evolution.} 
We will return to this in more detail in a future work where we plan to extend the 2PI analysis beyond the Hartree level and quantify the magnitude of off-shell corrections by comparing the results against classical dynamics.   

\begin{figure}[t]
     \centering
     \includegraphics[width=0.95\linewidth]{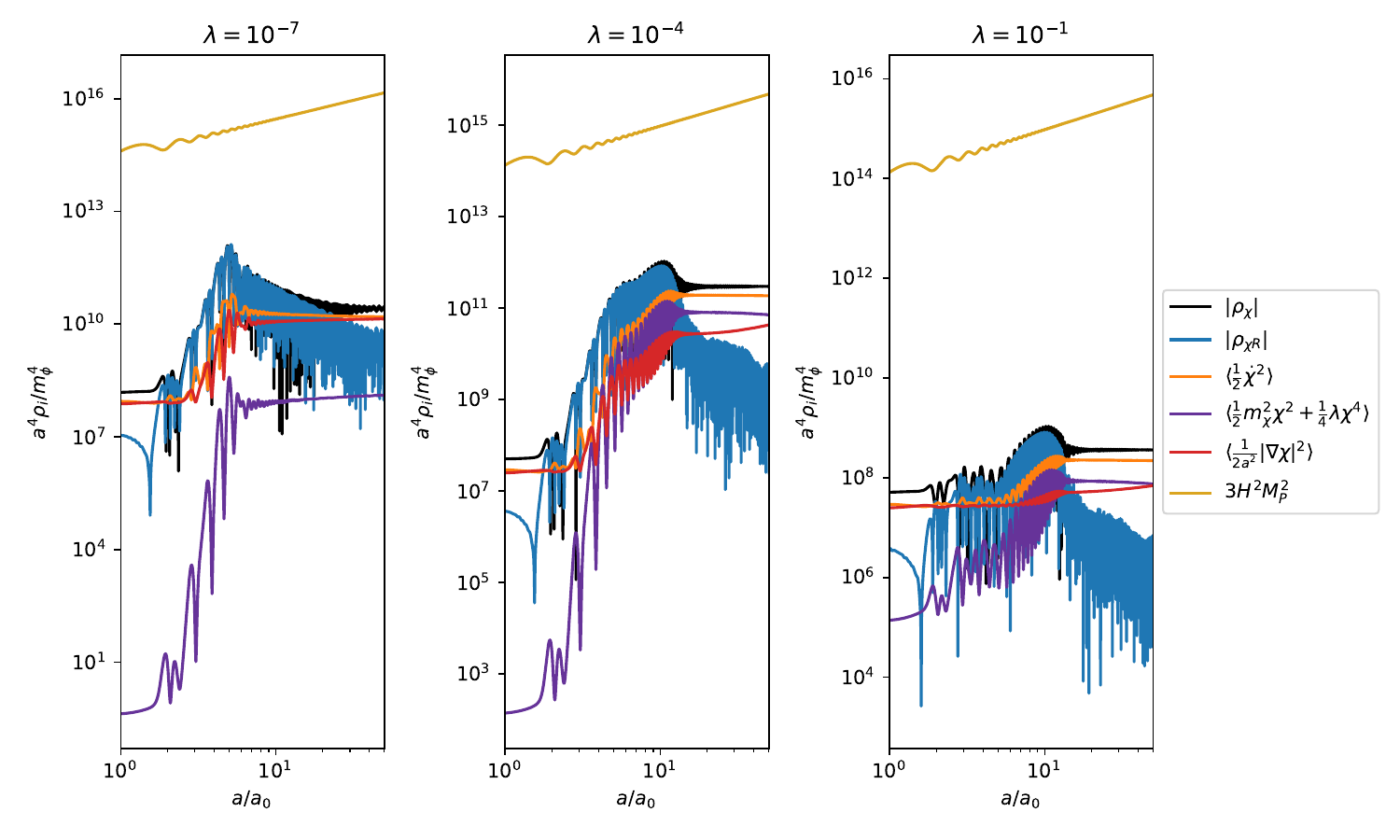} 
        \caption{The spectator field energy density and its components for runs with $\lambda = 10^{-7}$, $\lambda = 10^{-4}$ and $\lambda = 10^{-1}$ compared with the total inflaton energy density. The interaction energy of the non-minimal coupling is defined as $\rho_{\chi R} = \langle 3 \xi(H^2 \chi^2 + 2 H \chi\dot{\chi} -\frac{2}{3 a^2}\nabla\cdot(\chi\nabla\chi))\rangle$.}
        \label{fig:rhoplot}
\end{figure}

Finally, figure \ref{fig:rhoplot} shows the spectator energy densities (\ref{eq:rho_chi}) measured from the lattice solutions. It can be seen that in all cases the spectator energy density remains clearly subdominant compared to the total energy density, as we have assumed from the outset of the analysis. Note that this is different from the lattice simulation performed in \cite{Figueroa:2021iwm} for a similar non-minimally coupled scalar but with a different reheating equation of state which caused the initial spectator to grow to a dominant energy component. 

\begin{figure}[t]
     \centering
     \includegraphics[width=0.95\linewidth]{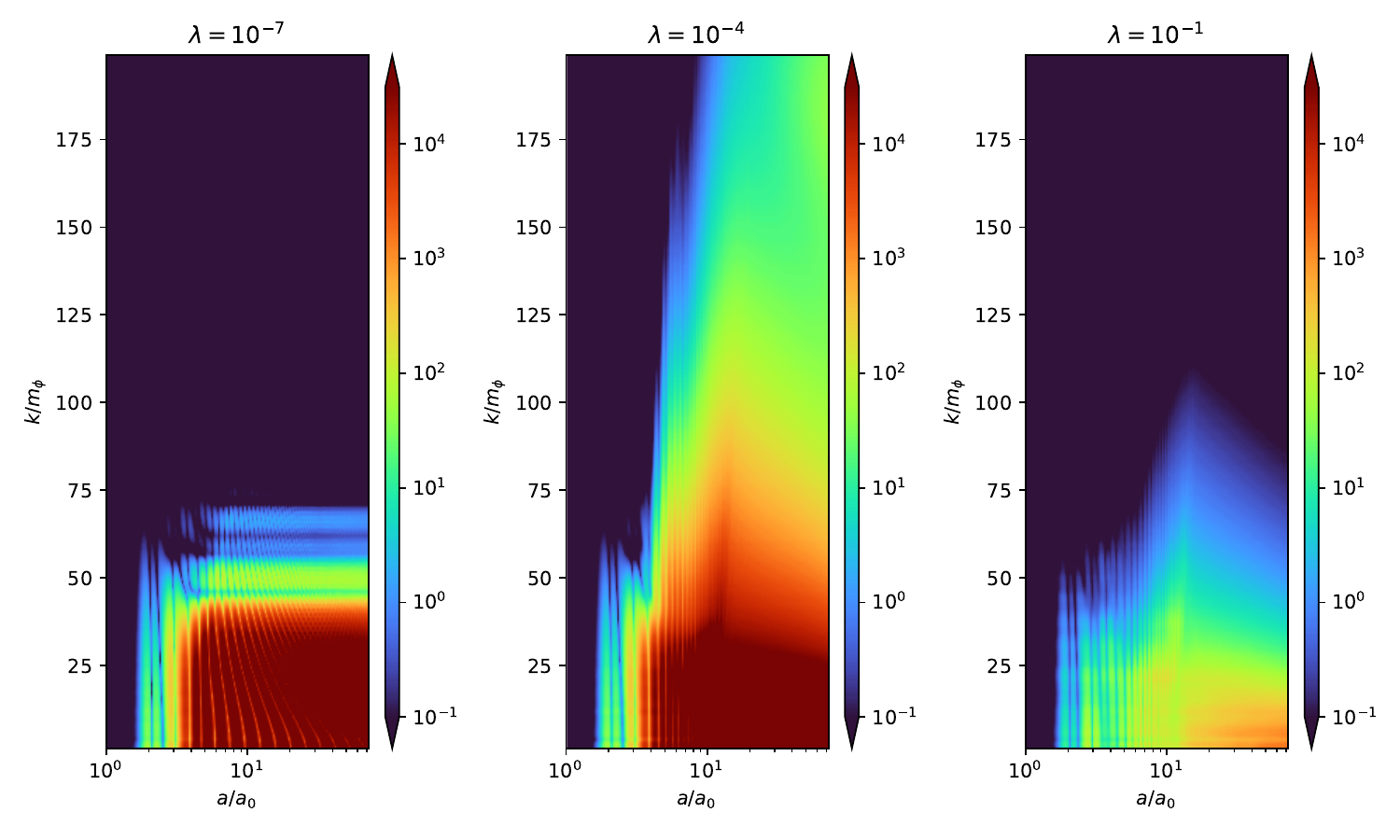} 
        \caption{The vacuum-subtracted lattice power spectra $a^2 (P(k) - P_{\rm vac}(k)) / m_{\phi}^{-1}$ for runs with $\lambda = 10^{-1}$, $\lambda = 10^{-4}$ and $\lambda = 10^{-7}$. }
        \label{fig:spectra_three_runs}
\end{figure}

\paragraph{Spectra.} 

\begin{figure}[t]
     \centering
     \includegraphics[width=0.95\linewidth]{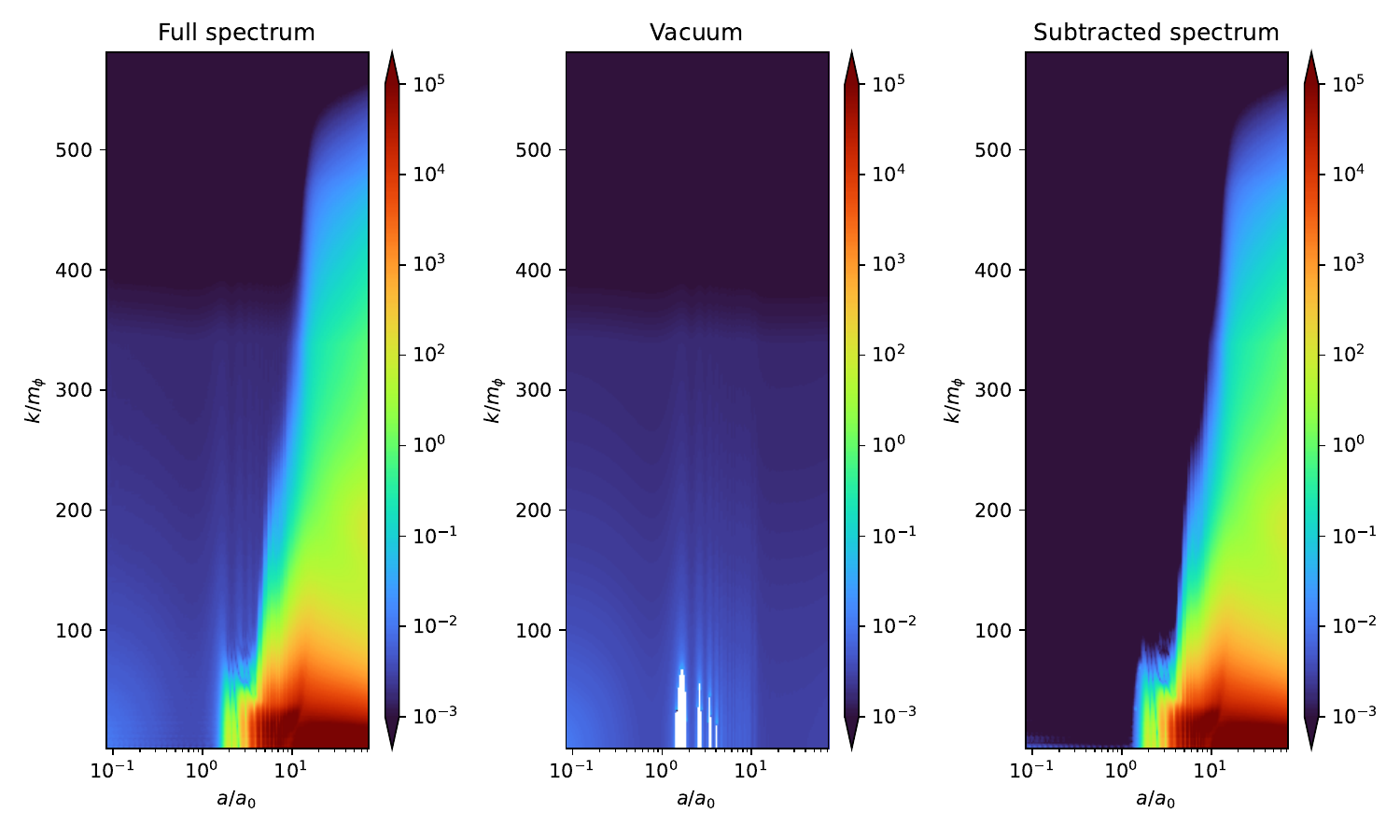} 
        \caption{The measured lattice power spectrum $a^2 P(k)/m_{\phi}^{-1}$ (left panel), the vacuum power spectrum calculated using the effective mass $a^2 P_{\rm vac}(k)/m_{\phi}^{-1}$(middle panel) and the vacuum-subtracted power spectrum $a^2 (P(k) - P_{\rm vac}(k)) / m_{\phi}^{-1}$ (right panel) for a run with $\lambda = 10^{-4}$. The unstable modes are denoted with white in the middle panel. Note that they overlap neatly with the initial tachyonic growth in the middle panel of figure~\cref{fig:spectra_three_runs}. }
        \label{fig:spectra_vacuum_subtraction}
\end{figure}

The vacuum subtracted power spectra $P(k)-P_{\rm vac}(k)$ measured from the lattice solutions are shown in figure~\cref{fig:spectra_three_runs}. Here $P(k)$ denotes the power spectrum of the full two point function, $\langle \chi({\bf k})\chi({\bf k'}) \rangle = (2\pi)^3 \delta({\bf k}+{\bf k'}) P(k)$, and the subtracted vaccum part $P_{\rm vac}(k)$ is given by equation (\ref{eq:lattice_vacuum}). The subtraction procedure is illustrated in figure~\cref{fig:spectra_vacuum_subtraction} which shows $P(k)$, $P_{\rm vac}(k)$ and $P(k)-P_{\rm vac}(k)$ for $\lambda = 10^{-4}$. 

The initial tachyonic growth of the power spectra is similar in all three cases but their subsequent evolution differs considerably from each other. For $\lambda = 10^{-7}$ the dynamics remains essentially linear after the tachyonic stage, and the horizontal bands seen in the left panel of figure~\cref{fig:spectra_three_runs} arise from the time dependent effective mass term $\xi R \chi^2$ \cite{Bassett:1997az}. For $\lambda=10^{-4}$ and $10^{-1}$, the dynamics becomes strongly non-linear after the tachyonic stage and the self-interactions spread the power rapidly towards higher momenta. This effect is weaker in the $\lambda = 10^{-1}$ -case, as there is less particle production overall. In the $\lambda = 10^{-4}$ -case, however, the entire momentum range of the lattice is populated. The drive towards larger momenta occurs most prominently for $a/a_0 \approx 10$, which coincides with the resonant growth of the two-point function. Afterwards, a general flow of power back towards lower momenta can be seen in both runs. An exception to this trend is the appearance of a secondary peak in the $\lambda = 10^{-4}$ -spectrum at $k/m_{\phi} \approx  200$, better visible in figure~\cref{fig:spectra_vacuum_subtraction}. We will discuss this feature in more detail later. 

For the nearly free-field case $\lambda = 10^{-7}$ the lattice spectrum is essentially identical with the 2PI Hartree result of~\cite{Kainulainen_2023}. This is expected and provides a consistency check for the lattice implementation. For the non-trivial cases $\lambda = 10^{-4}$ and $\lambda = 10^{-1}$, the lattice spectra differ from the results of~\cite{Kainulainen_2023} considerably more than the variances. The spectra start to deviate from the 2PI Hartree results once the non-linear evolution begins. On the lattice, the transfer of power towards larger $k$ begins slightly earlier, takes a longer time and the spectra never develop the clean UV cutoffs seen in~\cite{Kainulainen_2023}. In addition, the band structures seen in~\cite{Kainulainen_2023} and associated to non-linear resonance driven by the two-point function are not visible in the lattice results but this is likely due to limitations of the numerical resolution. Note that the bands seen for $\lambda = 10^{-7}$ both in the lattice and 2PI solutions are due to oscillations of the background field $R(t)$ and therefore not directly linked to the resolution with which the non-linear evolution is probed.   

\begin{figure}[t]
     \centering
     \includegraphics[width=0.95\linewidth]{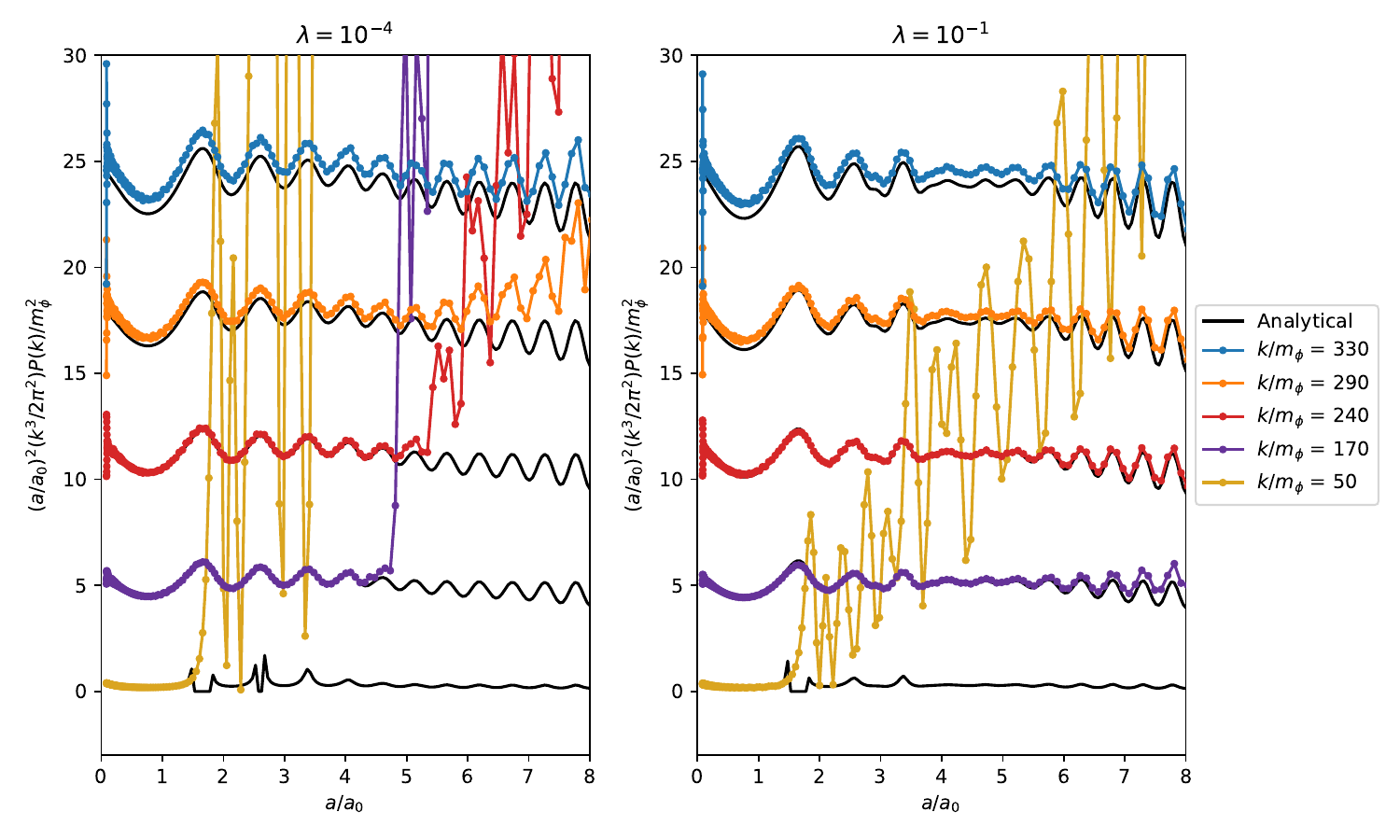} 
        \caption{A comparison between the measured lattice power spectrum and the analytical vacuum power calculated using the effective mass solved from equation \cref{eq:gap_equation} at various momenta for runs with $\lambda = 10^{-4}$ (left panel) and $\lambda = 10^{-1}$ (right panel). The runs were performed with $N = 512$, which corresponds to a linear UV cutoff at $k/m_\phi \approx 337$.}
        \label{fig:spectra_vacuum_evolution}
\end{figure}

Finally, figure~\cref{fig:spectra_vacuum_evolution} depicts the time evolution of the full lattice spectrum $P(k)$ and the vacuum spectrum $P_{\rm vac}(k)$ given by equation (\ref{eq:lattice_vacuum}). For each $k$-value in the figure, $P(k)$ initially closely tracks $P_{\rm vac}(k)$, whose time evolution is determined by the effective mass solved from the gap equation (\ref{eq:gap_equation}). This confirms that the gap equation, which a priori has no dynamical relevance whatsoever in the lattice computation, properly determines the effective mass of the system and justifies our definition of the vacuum spectrum (\ref{eq:lattice_vacuum}). Note that this is a non-trivial check, as especially the right panel with $\lambda= 10^{-1}$ shows modes for which $P(k)$ tracks $P_{\rm vac}(k)$ deep into the non-linear region, where the effective mass is dominated by the two-point function, see the left panel of figure~\ref{fig:variances_M2eff}. The full spectrum $P(k)$ starts to deviate from $P_{\rm vac}(k)$ once the mode $k$ undergoes a tachyonic instability, ($k/m_{\phi} =50$ curves in the figure), or particle production due to the non-linear resonance, (other curves in the figure). For the largest-$k$ -modes the lattice resolution starts to become insufficient, and the lattice spectrum shows a small constant shift away from $P_{\rm vac}(k)$. 

\paragraph{Discussion on the numerical resolution.} 

As mentioned above, after the non-linear resonance, the variances measured from the lattice for $\lambda = 10^{-4}$ and $\lambda = 10^{-1}$ scale nearly proportional to $a^{-2}$, in agreement with the 2PI Hatree results of~\cite{Kainulainen_2023} and as expected in the absence of particle production. This continues until $a/a_0 \sim 30$ after which the variances start to decrease slightly faster. At least part of this deviation from the asymptotic $a^{-2}$-scaling appears to be a numerical artefact related to the lattice resolution\footnote{For $\lambda = 10^{-7}$ the variance instead decreases slightly slower than $a^{-2}$, see figure~\ref{fig:varentsiances_M2eff}. This is a true physical effect due to slow particle production sourced by the time dependent mass term $\xi R \chi^2$, and in agreement with~\cite{Kainulainen_2023}.}. 
\begin{figure}[t]
     \centering
     \includegraphics[width=0.95\linewidth]{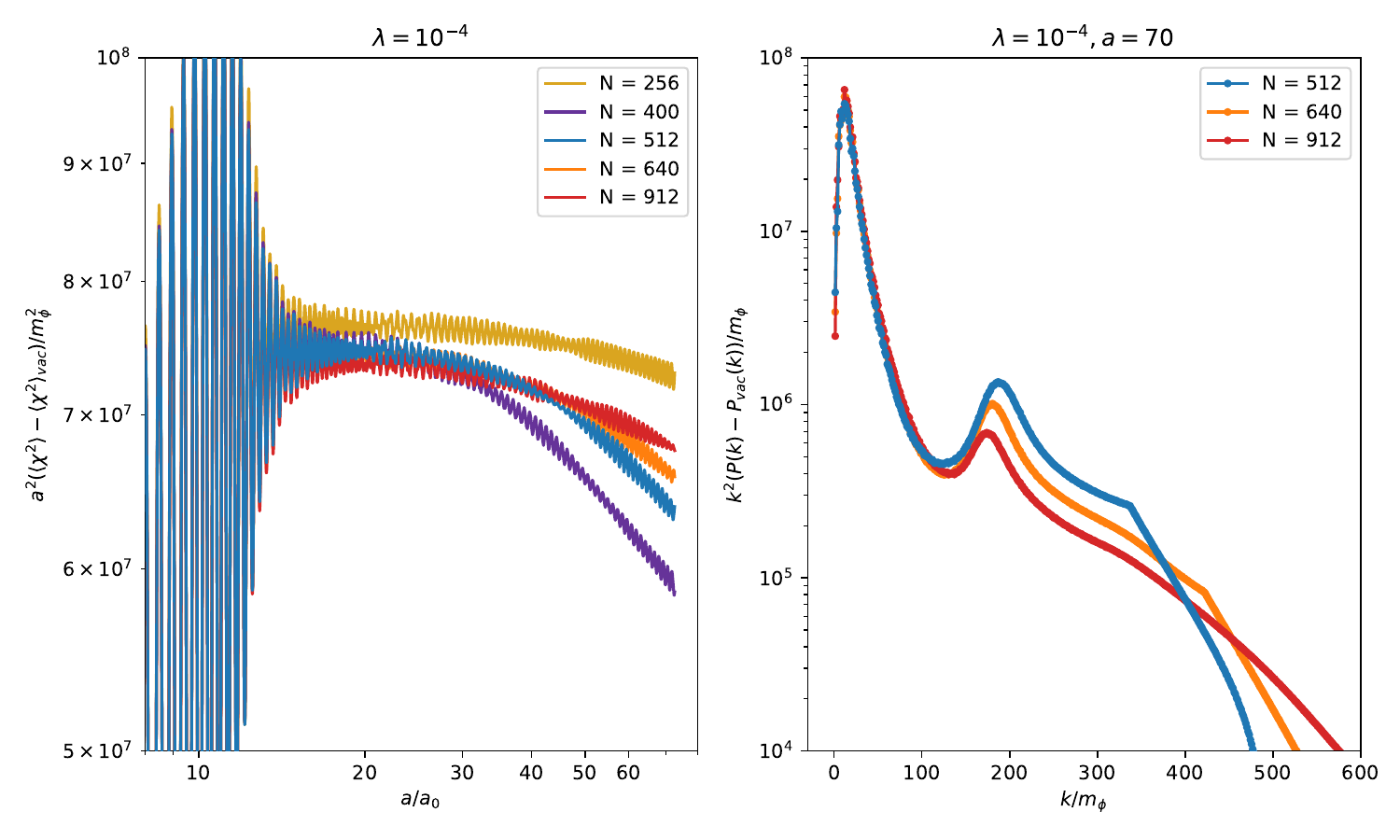} 
        \caption{(Left panel.) Late evolution of the spectator field variance calculated with various lattice grid sizes $N$ for $\lambda = 10^{-4}$. (Right panel.) The corresponding lattice power spectra measured at $a = 70$ with different $N$.}
        \label{fig:variance_resolution_dependence}
\end{figure}
This is elaborated in figure~\cref{fig:variance_resolution_dependence} where the left panel shows the variances for $\lambda = 10^{-4}$ obtained using different resolutions. For the lowest resolution shown in the figure, $N = 256$, the asymptotic scaling of the variance is closest to $a^{-2}$. Increasing the resolution to $N = 400$ generates a clear deviation from the $a^{-2}$ scaling and also slightly changes the solution at the end of the resonance. When the resolution is further increased, the scaling again gradually approaches the $a^{-2}$ behaviour. For $N \geqslant 400$, the results neatly overlap with each other until $a/a_0 \approx 30$. The higher resolution results therefore appear to be reliable up to this point. We also note that increasing the physical lattice box size in the same proportion with $N$, so as to not change the lattice UV cutoff, leaves the solutions unchanged. Also, the solutions for $\lambda = 10^{-7}$ agree with each other for all resolutions tested.    

There is a similar resolution-dependent effect in the spectra. The right panel of figure  ~\cref{fig:variance_resolution_dependence} shows the spectra for $\lambda = 10^{-4}$ obtained using different resolutions. As the resolution is increased, the spectra begin to develop a secondary peak about $k/m_{\phi} \approx 200$. The location of the peak is nearly fixed between different resolutions but we have checked that changing $\lambda$ moves the peak somewhat. As $N$ is increased, the peak first becomes more pronounced, but for large $N$ it begins to disappear. At the same time the power near the UV cutoff decreases with the increasing resolution.

The deviation of the variance from the asymptotic $a^{-2}$ scaling is driven by a decrease in power for $k$-modes between the primary IR peak and the secondary peak in figure~\cref{fig:variance_resolution_dependence}. The secondary resolution-dependent peak appears to enhance the transfer of power from the primary peak to higher momenta, where the power can leak beyond the UV cutoff of the lattice simulation. It then appears, that the deviation of the variances from the scaling $a^{-2}$ for $a/a_0 \gtrsim 30$ is at least partially a resolution dependent spurious effect. 

%
\section{Conclusions and discussion}
\label{sec:conclusions}
%

We have performed classical lattice simulations to study the out-of-equilibrium dynamics during reheating for a non-minimally coupled self-interacting spectator scalar $\chi$ which contributes to dark matter \cite{Markkanen:2015xuw,Fairbairn:2018bsw}.  The quantum evolution in  the same scenario was previously investigated in~\cite{Kainulainen_2023} using the non-perturbative 2PI approach in the Hartree truncation. The dark matter production in the setup \cite{Markkanen:2015xuw,Fairbairn:2018bsw} is initially driven by the non-minimal coupling $\xi R\chi^2$ which leads to a tachyonic instability when the Ricci scalar is oscillating from positive to negative values during reheating.  In~\cite{Kainulainen_2023}  it was found that the tachyonic stage is followed by a novel non-linear resonance driven by the two-point function $\langle \chi^2 \rangle$ in the presence of the self-interaction $\lambda \chi^4$, and the resonance can dominate the net particle yield. However, the results of~\cite{Kainulainen_2023} are obtained in the Hartree approximation, which neglects momentum exchanging interactions. 

In this work we have made a detailed comparison between the 2PI Hartee results of~\cite{Kainulainen_2023} and a classical lattice implementation of the setup using the CosmoLattice code~\cite{Figueroa_2021,Figueroa_2023}. For a small self-coupling $\lambda = 10^{-7}$ we find that the lattice results coincide with the 2PI Hartee results through the entire simulation run. This is expected as in this case the self-interactions remain negligible and the dynamics is essentially linear. The Hartree approximation used in~\cite{Kainulainen_2023} then becomes exact, and in the linear case there should also be no differences between the quantum and classical evolution of the two point function. The lattice solution for the initial near-linear tachyonic phase is in agreement with the 2PI Hartree results for $\lambda=10^{-4}$ and $\lambda=10^{-1}$ as well. Importantly, the lattice simulations in these cases show a transient stage of non-linear resonance at the end of the tachyonic period in agreement with the findings of~\cite{Kainulainen_2023}. The final amplitudes of $\langle \chi^2\rangle$ measured from the lattice agree up to a factor of $\mathcal{O} (1)$ with those of~\cite{Kainulainen_2023}. However, the transient resonance is stronger on the lattice than in the 2PI Hartree calculation, leading to larger final amplitudes, contrary to what one might expect.  The most significant differences between the two approaches are seen in the spectrum of the two-point function. On the lattice, the power spectrum extends to considerably larger momenta during the non-linear part of the evolution than in the Hartree-solutions, which shows a clear UV cutoff instead. We expect the differences are mostly due to momentum exchanging processes mediated by the scalar self-interactions, which the Hartree truncation does not account for. 

Overall, our results show that detailed resolution of non-linear out-of-equilibrium processes can be necessary to determine the dark matter abundance and momentum distribution in setups with primordial particle production. In~\cite{Markkanen:2015xuw,Fairbairn:2018bsw}, the $\mathcal{O} (1)$ changes in the scalar two-point function induce comparable changes in the final dark matter abundance, see~\cite{Fairbairn:2018bsw} for details. The dark matter momentum distribution, which affects the structure formation, depends on the spectrum of the two-point function, which we find to be very sensitive to details of the non-linear evolution. While quantum corrections are generally suppressed in the limit of large occupation numbers, it is not clear what their quantitative is effect in scenarios like~\cite{Markkanen:2015xuw,Fairbairn:2018bsw}. Such setups involve coherent fields and strongly non-linear evolution, which cannot be perturbatively modelled in terms of interacting asymptotic particle states. It is therefore important to develop computationally feasible methods to study the primordial out-of-equilibrium evolution at quantum level, including all dynamically relevant interactions. 

We will return to this in a future work where we plan to extend the 2PI analysis of~\cite{Kainulainen_2023} beyond the Hartree level by including collision terms in the coupled equations of quantum one- and two-points functions similar to~\cite{Kainulainen:2021eki}. Using this approach, we expect to be able to quantify the importance quantum off-shell corrections compared to the on-shell dynamics captured by the classical lattice simulations performed in this work. In addition, it can be noted that while the computation time of the lattice simulation, (at least for a plain implementation in direct space), scales proportional to $N^3$, the 2PI Hartree computation of~\cite{Kainulainen_2023} instead scales linearly proportional to the number of $k$-modes. It will be interesting to see whether the collision integrals required beyond the Hartree approximation can be implemented such that part of this advantage can be retained.

%
\section*{Acknowledgements}
\label{sec:ack}
We thank Dani Figueroa for helpful discussions on the CosmoLattice-package. The work of OV was supported by a grant from the Magnus Ehrnrooth foundation.
%

%
\bibliography{maindesc.bib}
%

%
\end{document}